\documentclass{birkjour}
\usepackage{cite}
 \newtheorem{thm}{Theorem} 

\newtheorem{conjecture}[thm]{Conjecture}
 \theoremstyle{definition}
 \newtheorem{defn}[thm]{Definition}
 \theoremstyle{remark}

\begin{document}

%
%
%
%
%
%
%
%
%

\title [Questioning the Standard Model with the Deformation Philosophy]
{``The important thing is not to stop \\ questioning", including
the symmetries \\ on which is based the Standard Model} 
\thanks{This text is published in
Geometric Methods in Physics, XXXII Workshop 2013 in 
Bia{\l}owie{\.z}a, Trends in Mathematics, 7-37, Springer (2014).  \\ 
\copyright{\footnotesize 2014 Daniel Sternheimer}}

\author[Daniel Sternheimer]{Daniel Sternheimer}

\address{%
Department of Mathematics, Rikkyo University, \\
3-34-1 Nishi-Ikebukuro, Toshima-ku, Tokyo 171-8501, Japan,\\
\& Institut de Math\'ematiques de Bourgogne, Universit\'e de Bourgogne\\
BP 47870, F-21078 Dijon Cedex, France. \\ 
}

\email{Daniel.Sternheimer@u-bourgogne.fr}




\date{January 1, 2014}
\dedicatory{To the memory of Moshe Flato and of Noriko Sakurai} 
\vspace*{-9mm}


\begin{abstract}
New fundamental physical theories can, so far a posteriori, be seen as 
emerging from existing ones via some kind of deformation. That is the
basis for Flato's ``deformation philosophy", of which the main paradigms are
the physics revolutions from the beginning of the twentieth century,
quantum mechanics (via deformation quantization) and special relativity. 
On the basis of these facts we describe two main directions by which 
symmetries of hadrons (strongly interacting elementary particles) may ``emerge" 
by deforming in some sense (including quantization) the Anti de Sitter symmetry 
(AdS), itself a deformation of the Poincar\'e group of special relativity. 
The ultimate goal is to base on fundamental principles the dynamics of strong 
interactions, which originated half a century ago from empirically guessed 
``internal" symmetries. After a rapid presentation of the physical (hadrons)
and mathematical (deformation theory) contexts, 
we review a possible explanation of photons as composites of AdS singletons 
(in a way compatible with QED) and of leptons as similar composites 
(massified by 5 Higgs, extending the electroweak model to 3 generations). 
Then we present a ``model generating" multifaceted framework in which AdS 
would be deformed and quantized (possibly at root of unity and/or in manner 
not yet mathematically developed with noncommutative ``parameters"). 
That would give (using deformations) a space-time origin to the 
``internal" symmetries of elementary particles, 
on which their dynamics were based, and either question, or give a 
conceptually solid base to, the Standard Model, in line with Einstein's 
quotation: \textsl{``The important thing is not to stop questioning. 
Curiosity has its own reason for existing."}   
\end{abstract}

\maketitle

\section {Introduction: the deformation philosophy and the present proposal}

\subsection{Why deformations?}
However seductive the idea may be, the notion of ``Theory of Everything" 
is to me unrealistic. In physics, knowingly or not, one makes approximations 
in order to have as manageable a theory (or model) as possible. That happens 
in particular when the aim is to describe the reality known at the time,
even if one suspects that a more elaborate reality is yet to be discovered. 
The question is how to discover that reality. We claim, on the basis of 
past experience, that one should not extrapolate but rather ``deform."
   
Indeed physical theories have their domain of applicability defined, e.g., by 
the relevant distances, velocities, energies, etc. involved. But the passages 
from one domain (of distances, etc.) to another do not happen in an 
uncontrolled way: experimental phenomena appear that cause a paradox and 
contradict accepted theories, in line with the famous quote by Fermi 
\cite{Fermi1}: \textsl{There are two possible outcomes: if the result confirms 
the hypothesis, then you've made a measurement. If the result is contrary to 
the hypothesis, then you've made a discovery.} 

Eventually a new fundamental constant enters, causing the formalism to be 
modified: the attached structures (symmetries, observables, states, etc.) 
\textsl{deform} the initial structure to a new structure which in the limit, 
when the new parameter goes to zero, ``contracts" to the previous formalism.
The problem is that (at least until there is no other way out) the physics 
community is gregarious. Singing ``It ain't necessarily so" (which I am doing 
in this paper) is not well received.  

A first example of the ``deformation" phenomenon can be traced back to  
Antiquity, when it was gradually realized that the earth is not flat. 
[Yet nowadays some still dispute the fact!] In mathematics the first instances 
of deformations can be traced to the nineteenth century with Riemann surface 
theory, though the main developments happened a century later, in particular 
with the seminal analytic geometry works of Kodaira and Spencer \cite{KS58} 
(and their lesser known interpretation by Grothendieck \cite{Gr61}, where one 
can see in watermark his ``EGA" that started a couple of years later). These 
deep geometric works were in some sense ``linearized" in the theory of 
deformations of algebras by Gerstenhaber \cite{Ge64}. 

The realization that deformations are fundamental in the development of 
physics happened a couple of years later in France, when it was noticed that                                                     
the Galilean invariance 
($SO(3)\cdot\mathbb{R}^3\cdot\mathbb{R}^4$) of Newtonian mechanics is deformed, 
in the Gerstenhaber sense \cite{Ge64}, to the Poincar\'e group of special         
relativity ($SO(3,1)\cdot\mathbb{R}^4$). In spite of the fact 
that the composition law of symbols of pseudodifferential operators, essential
in the Atiyah--Singer index theorem developed at that time (to the exposition 
of which I took part in Paris in the S\'eminaire Cartan--Schwartz 1963/64),       
was in effect a deformation of their abelian product, it took another             
ten years or so to develop the tools which enabled us to make explicit, 
rigorous and convincing, what was in the back of 
the mind of many: quantum mechanics is a deformation of classical mechanics. 
That developed into what became known as \emph{deformation quantization} 
and its manifold avatars and more generally into the realization that 
quantization is deformation. This stumbling block being removed, 
the paramount importance of deformations in theoretical physics became
clear \cite{Fl82}, giving rise to what I call ``Flato's deformation philosophy".

This paper being aimed at both physicists and mathematicians and dealing with    
so many topics, we may look overly schematic (even trivial) in many parts 
for readers coming from one or the other community. More details can be found 
in relatively recent reviews (\cite{DS01,St07,St12} by myself, and many more 
by others) and references quoted therein.
The hope is that both communities will get the flavor of (and maybe contribute 
to) the framework for models proposed here. It is based on developments I have 
witnessed since the early 1960s and in which Flato and I took part, sometimes 
in a controversial way. In numerous discussions I had with scientists 
around these ideas, especially in the past two years, I was surprised to 
notice that many had often only a vague idea of a number of the topics 
involved, not going beyond the views given in textbooks and/or educated 
popularizations. I am probably one of the very few who can 
(and dare) deal with all the topics involved in that unconventional manner. 
Part of the mathematical aspect is virgin territory and in any case requires an 
approach (which can be called ``mathematical engineering") dealing more 
with specific examples than with very abstract developments. On the other hand
the basic physical approach is unconventional, and some of the physical issues 
and models questioned here have for years been taught as facts in courses 
and presented as such in the literature.   
  
\subsection{A brief overlook of the paper}
Towards the end of the nineteenth century many believed that, in particular 
with Newtonian mechanics (and gravitation) and electromagnetism, physics was 
well understood. Yet the best was to come. In the first half of last century 
appeared  relativity and quantum mechanics, which we now can interpret as 
deformations. On the fundamental side the second half of last century was 
dominated by the interactions between elementary particles, classified 
(in increasing order of strength) as gravitational, weak, electromagnetic and 
strong. Quantum electrodynamics (QED), developed in the 1940s, explained 
electromagnetic interactions with an extremely  high level of accuracy (even 
if the theory is not yet fully mathematically rigorous). In the 1970s it was
combined with weak interactions in the electroweak model, which required the 
Higgs boson that was (most likely) now discovered in CERN. 

After an outlook of the physical and mathematical context we shall indicate how, 
using AdS symmetry (a deformation of Poincar\'e) we can explain the photon (the 
basis of QED) as composite of two ``singletons", massless particles in 
a 2+1 space-time (themselves composites of two harmonic oscillators). Then 
an extension of the electroweak model to the presently known 3 generations 
of leptons could explain how, in AdS, these can also be composites of
singletons, massified by 5 Higgs. 

It is therefore tempting to try and obtain the symmetries of hadrons, 
on which their dynamics has been built, by deforming further AdS. That cannot 
be done in the category of Lie groups but can, e.g., in that of Hopf algebras 
(quantum groups). It turns out that these, at root of unity (often called 
``restricted quantum groups") are finite dimensional vector spaces, and have 
finite dimensional UIRs (unitary irreducible representations), an important 
feature of the presently used simple unitary symmetries. There are of course 
many other problems to address, which cannot be ignored, but if that direction 
produces a model which could fit experimental data, a revolution in our 
understanding of physics might follow.  

That could be too much to hope for and more general deformations might be 
needed, in particular (also at roots of unity) multiparameter (e.g., 
with parameters in the group algebra of ${\mathbb{Z}}/{n\mathbb{Z}}$, 
denoted in the following by $\mathbb{Z}_{(n)}$), or a novel theory of deformations, 
not yet developed mathematically, with non-commutative ``deformation parameter" 
(especially quaternions or belonging to the group algebra of $\mathbb{S}_n$, 
the permutation group of $n$ elements, e.g., $n=3$).   

Both are largely virgin mathematical territory, and if successful we might 
have to ``go back to the drawing board," for the theory and for the 
interpretation of many raw experimental data. That is a challenge worthy 
of the future generations, which in any case should give nontrivial mathematics.

\section{A very schematic glimpse on the context: hadrons and their symmetries}
In the fifties the number of known elementary particles increased so 
dramatically that Fermi quipped one day \cite{Fermi1}: \textsl{Young man, if 
I could remember the names of these particles, I would have been a botanist.}

Clearly, already then, the theoretical need was felt, to bring some order into
that fast increasing \cite{PDG13} flurry of particles. Two (related) natural 
ideas appeared: To apply in particle physics ``spectroscopy" methods 
that were successful in  molecular spectroscopy, in particular group 
theory \cite{Wi59}. And to try and treat some particles as ``more elementary", 
considering others as composite. 

A seldom mentioned caveat: In molecular spectroscopy, e.g., when a 
crystalline structure breaks rotational symmetry, which (for trigonal and
tetragonal crystals) was the subject of Flato's M.Sc. Thesis \cite{KiblerCMF} 
under Racah (defended in 1960 and still frontier when its main part was 
published in 1965 as his French ``second thesis"), we know the forces, and 
their symmetries give the spectra (energy levels). In particle physics things 
occurred in reverse order: one guessed symmetries from the observed spectra, 
interpreted experimental data on that basis and developed dynamics 
compatible with them. 

In the beginning, in order to explain the similar behavior of proton $p$ and 
neutron $n$ under strong interactions, a quantum number (isospin) was introduced 
in the 1930s, related to a $SU(2)$ symmetry. In the 1950s new particles were 
discovered in cosmic rays, that behaved ``strangely" (e.g., they lived much 
longer than expected). So a new quantum number (strangeness) was introduced, 
which would be conserved by strong and electromagnetic interactions, but not 
by weak interactions. One of these is the baryon $\Lambda$. In~1956 Shoichi
Sakata \cite{Sa56}, extending an earlier proposal by Fermi and Yang (involving 
only protons and neutrons) came with the ``Sakata model"  according to which 
$p$, $n$ and $\Lambda$ are ``more elementary" and the other particles are 
composites of these 3 and their antiparticles. This conceptually appealing 
model (maybe not as ``sexy" as Yoko Sakata, a top model who was not born then) 
had a strong impact \cite{Ok07}, in spite of the fact that a number of the 
experimental predictions it gave turned out to be wrong. 

In the beginning of 1961 an idea (that was in the making before) appeared: 
since we have 2 quantum numbers (isospin and strangeness, we would now say
``two generations") conserved in strong interactions, we should try a rank 2 
compact Lie group to ``put into nice boxes" the many particles we had.
In particular three papers were written then: An elaborate paper \cite{BDFL62} 
in which, ``since it is as yet too early to establish a definite 
symmetry of the strong interactions," all 3 groups (types $A_2$, $B_2 = C_2$ 
and $G_2$), and more, were systematically studied. And two \cite{GM61,Ne61}, 
in which only the simplest ($SU(3)$, type $A_2$) was proposed. 
The known ``octets" of 8 baryons of spin $\frac{1}{2}$ and of the 
8 scalar (spin 0) and vector (spin 1) mesons fitted nicely in the 
8-dimensional adjoint representation. (Hence the name ``the eightfold way" 
coined by Gell-Mann, an allusion to the ``Noble Eightfold Path of Buddhism".) 
In 1962 Lev Okun proposed ``hadrons" as a common name for strongly interacting 
particles, the half-integer spin (fermions) baryons, usually heavier, and the
integer spin (bosons) mesons. The 9 then known baryons of spin~$\frac{3}{2}$
($4\ \Delta +3\ \Sigma^* +2\ \Xi^*$) were associated with the 10 dimensional 
representation: the missing one (${\Omega}^{-}$) in the ``decuplet" was 
discovered in 1964 with roughly the properties predicted by Gell-Mann in 1962. 
Big success! (Even if anyone can guess that after 4,3,2 comes 1...)

But what to do with the basic (3-dimensional) representations of $SU(3)$, 
which can give (by tensor product and reduction into irreducible components) 
all other representations? In 1964 Murray Gell-Mann, and independently George 
Zweig, suggested that they could be associated with 3 entities (the same 
number as in the Sakata model) and their antiparticles. Zweig proposed to 
call them ``aces" but Gell-Mann, with his feeling for a popular name, 
called them ``quarks", a nonsense word which he imagined and shortly afterward
found was used by James Joyce in ``Finnegans Wake:"
\begin{center}
\textsl {Three quarks for Muster Mark!\\
Sure he has not got much of a bark \\
And sure any he has it's all beside the mark.}
\end{center}
Now, how could such ``confined" quarks, which would have spin $\frac{1}{2}$ 
(not to mention fractional charge), coexist in a hadron, something forbidden 
by the Pauli exclusion principle? That same year O.W. Greenberg (and 
Y. Nambu) proposed to give them different ``colors", now labeled blue, green, 
and red. Eventually, since the 1970s, that gave rise to QCD (quantum 
chromodynamics) in parallel with QED but with nonabelian ``gauge group" $SU(3)$ 
instead of the abelian group $U(1)$ in QED. In order to keep them together 
``gluons" were introduced, which carried the strong force. From that time on, 
the development of particle physics followed essentially a ballistic 
trajectory, and eventually its theory became more and more 
phenomenology-oriented --~with the caveat that many
raw experimental data are interpreted within the prevalent models.    
 
In 1964 quarks came in 3 ``flavors" (up, down, and strange) but the 
same year a number of people, in particular Sheldon Glashow, proposed a 
fourth flavor (named charm) for a variety of reasons, which became gradually 
more convincing until in 1974  a ``charmed" meson J/$\Psi$ was discovered,
completing the 2 generations of quarks, in parallel with the 2 generations 
of leptons ($e$ and $\mu$) and their associated neutrinos. 
The number of supposed quark flavors grew to the current six in 1973, 
when Makoto Kobayashi and Toshihide Maskawa noted that an experimental 
observation ($CP$ violation) could be explained if there were another pair of 
quarks, eventually named bottom and top by Haim Harrari, and ``observed" 
(with much heavier mass
\footnote{The quark masses are not measurements, but parameters used in 
theoretical models and compatible with raw experimental data.} 
than expected for the top) at Fermilab in 1977 and 1995 (resp.). 
In parallel, in 1974-1977, the existence of a heavier lepton 
$\tau$ was experimentally found, and its neutrino discovered in 2000. Kobayashi
and Maskawa shared the 2008 Nobel prize in physics with Yoichiro Nambu who, 
already in 1960, described the mechanism of spontaneous symmetry breaking 
in particle physics. They were also awarded in 1985 the first J.J. Sakurai 
prize for Theoretical Particle Physics established, after JJ's premature 
death in 1982, with the American Physical Society (by his widow Noriko Sakurai, 
who in 2008 became my wife \cite{Fermi2}); Nambu had received the J.J. Sakurai 
prize in 1994. 

So now we have 3 generations of leptons and 3 of quarks (in 6 flavors 
and 3 colors). $SU(3)$ is back in, with a different meaning than originally. 
Eventually the electroweak model was incorporated and elaborate dynamics built 
on that basis of empirical origin, and everything seems to fit. 

In a series of recent papers (see \cite{CCS13} and references therein)
Alain Connes and coworkers showed that ``noncommutative geometry provides a 
promising framework for unification of all fundamental interactions 
including gravity." In the last paper, assuming that ``space-time is a 
noncommutative space formed as a product of a continuous four dimensional 
manifold times a finite space" he develops a quite personal attempt to predict 
the Standard Model (possibly with 4 colors). 

But what if the Standard Model was a colossus with clay feet (as in the
interpretation by prophet Daniel of Nebuchadnezzar's dream: Book of Daniel, 
Chapter 2, verses 31--36)? What if it were ``\textsl{all beside the mark}"?? 
    
\section{The mathematical context: Deformation theory and quantization}

In this section, for the sake of self-completeness, we shall give a very brief 
summary of what can be found with more details in a number of books, papers and
reviews (in particular \cite{DS01,St12}). Since quantization is a main paradigm
for our ``deformation philosophy", the idea is to give readers who would not 
know these already, some rudiments of deformation theory, of how quantum 
mechanics and field theory can be realized as a deformation of their classical 
counterparts, and of applications to symmetries (in particular the quantum 
group ``avatar"). Educated readers or those who do not care too much about
mathematical details may (at least for the time being \ldots) only browse 
through this Section. Note however that deformation quantization (as it is now 
known), introduced in the ``founding papers" \cite{BFFLS}, is more than a mere 
reformulation of usual quantum mechanics; in particular it goes beyond 
canonical quantization (on ${\mathbb R}^{2\ell}$) and applies to general phase 
spaces (symplectic or Poisson manifolds).   

\subsection{The Gerstenhaber theory of deformations of algebras}
A concise formulation of a Gerstenhaber deformation (over the field 
${\mathbb K}[[\nu]]$ of formal series in a parameter $\nu$ with coefficients 
in a field ${\mathbb K}$) of an algebra (associative, Lie, bialgebra, etc.) 
over ${\mathbb K}$ is \cite{Ge64,BFGP}:

\begin{defn}\label{DrGdef}
A deformation of an algebra $A$ over ${\mathbb K}$ is an algebra 
${\tilde A}$ 
over ${\mathbb K}[[\nu]]$ such that ${\tilde A}/\nu {\tilde A} \approx A$.
Two deformations ${\tilde A}$ and ${\tilde A'}$ are said equivalent if they
are isomorphic over ${\mathbb K}[[\nu]]$ and ${\tilde A}$ is said trivial if
it is isomorphic to the original algebra $A$ considered by base field
extension as a ${\mathbb K}[[\nu]]$-algebra.
\end{defn}

For associative (resp. Lie) algebras, the above definition tells us that 
there exists a new product $\ast$ (resp. bracket $[\cdot,\cdot]$) 
such that the new (deformed) algebra is again associative (resp. Lie).
Denoting the original composition laws by ordinary product 
(resp. Lie bracket $\{\cdot,\cdot\}$) this means, for $u_1,u_2 \in A$
(we can extend this to $A[[\nu]]$ by ${\mathbb K}[[\nu]]$-linearity), that we 
have the formal series expansions:
\begin{align}
u_1\ast u_2 &= u_1u_2 + \sum_{r=1}^\infty \nu^r C_r(u_1,u_2) \label{a}\\
\left[u_1, u_2\right] &= \{u_1,u_2\} + \sum_{r=1}^\infty \nu^r B_r(u_1,u_2)
\label{l}\end{align}
where the bilinear maps ($A\times A \to A$) $C_r$ and (skew-symmetric) $B_r$ 
are what are called 2-cochains in the respective cohomologies (Hochschild and
Chevalley--Eilenberg), satisfying (resp.) 
$(u_1\ast u_2) \ast u_3=u_1\ast (u_2\ast u_3) \in {\tilde A}$                 
and $\mathcal{S}[[u_1,u_2],u_3]=0$, for $u_1,u_2,u_3\in A$, $\mathcal{S}$ 
denoting summation over cyclic permutations, the leading term (resp. $C_1$ 
or $B_1$) being necessarily a 2-cocycle (the coefficient of $\nu$ in the
preceding conditions may be taken as a definition of that term). 

For a (topological) {\it bialgebra} (an associative algebra $A$ where we have
in addition a coproduct $\Delta : A \longrightarrow A \otimes A$ and the
obvious compatibility relations), denoting by $\otimes_\nu$ the tensor
product of ${\mathbb K}[[\nu]]$-modules, we can identify
${\tilde A}\, {\hat{\otimes}}_\nu {\tilde A}$ with
$(A\, {\hat{\otimes}}A)[[\nu]]$, where ${\hat{\otimes}}$ denotes the algebraic
tensor product completed with respect to some topology (e.g., projective 
for Fr\'echet nuclear topology on $A$). Then we have also a deformed
coproduct ${\tilde \Delta } = \Delta + \sum_{r=1}^\infty \nu^r D_r$,
$D_r \in \mathcal{L}(A, A {\hat{\otimes}}A)$ satisfying 
${\tilde \Delta }(u_1 * u_2)={\tilde \Delta }(u_1) * {\tilde \Delta }(u_2)$,
where $\mathcal{L}$ denotes the space of linear maps. 
In this context appropriate cohomologies can be introduced. There are 
natural additional requirements for Hopf algebras (see, e.g., \cite{BGGS}).

\subsection{ Deformation quantization}
The above abstract definition should become less abstract 
when applied to an algebra $N(W)$ of (differentiable) functions on a symplectic 
or Poisson manifold $W$, in particular those over phase space 
${\mathbb R}^{2\ell}$ (with coordinates $p, q \in {\mathbb R}^{\ell}$) endowed 
with the Poisson bracket $P$ of two functions $u_1$ and $u_2$, defined on a 
Poisson manifold $W$ as 
$P(u_1,u_2) = {\imath}(\Lambda)(du_1 \wedge du_2)$ (where ${\imath}$              
denotes the interior product, here of the 2-form $du_1 \wedge du_2$ with the 
2-tensor $\Lambda$ defining the Poisson structure on $W$, which in the case of 
a symplectic manifold is everywhere nonzero with for inverse a closed
nondegenerate 2-form $\omega$). For $W={\mathbb R}^{2\ell}$, $P$
can be written by setting $r=1$ in the formula for the $r^{\rm{th}}$ power 
($r\geq1$) of the bidifferential operator $P$ (we sum over repeated indices):
\begin{equation}  \label{PBr}    
P^r(u_1,u_2)=\Lambda^{i_1j_1}\ldots \Lambda^{i_rj_r}
(\partial_{i_1\ldots i_r}u_1)(\partial_{j_1\ldots j_r} u_2)
\end{equation}
with $i_k, j_k = 1,\ldots,2\ell$, $k=1,\ldots,r$
and $(\Lambda^{i_kj_k}) = {0\,-I\choose I\,0}$. We can write deformations
of the usual product of functions (deformations driven by the Poisson bracket) 
and of the Poisson bracket as what are now called the Moyal (``star") product 
and bracket, resp.
\begin{equation} \label{star}
u_1 \ast_M u_2 = \exp(\nu P)(u_1,u_2) = 
u_1u_2 + \sum^\infty_{r=1}\frac{\nu^{r}}{r!} P^{r}(u_1,u_2).
\end{equation}
\begin{equation}  \label{Moyal}
M(u_1,u_2) = \nu^{-1} \sinh(\nu P)(u_1,u_2) = P(u_1,u_2) + 
\sum^\infty_{r=1}\frac{\nu^{2r}}{(2r+1)!} P^{2r+1} (u_1,u_2). 
\end{equation}
These correspond (resp.) to the product and commutator of operators in the
``canonical" quantization on $\mathbb{R}^{2\ell}$  of a function $H(q,p)$ 
with inverse Fourier transform $\tilde{H}(\xi,\eta)$, given by (that formula
was found by Hermann Weyl \cite{We27} as early as 1927 when the weight is $\varpi=1$):
\begin{equation} \label{Weyl}
H \mapsto \hat{H}=\Omega_\varpi(H) = \int_{{\mathbb R}^{2\ell}} \tilde{H}(\xi,\eta)
{\exp}(i(\hat{p}.\xi + \hat{q}.\eta)/\hbar) \varpi(\xi,\eta) d^\ell \xi d^\ell \eta
\end{equation}
which maps the classical function $H$ into an operator on 
$L^2(\mathbb{R}^{2\ell})$, the ``kernel" 
${\exp}(i(\hat{p}.\xi + \hat{q}.\eta)/\hbar)$ being the corresponding unitary 
operator in the (projectively unique) representation of the Heisenberg group 
with generators $\hat{p_\alpha}$ and $\hat{q_\beta}$ 
($\alpha, \beta = 1, \ldots \ell$) satisfying the canonical commutation 
relations $[\hat{p_\alpha},\hat{q_\beta}] = i\hbar\delta_{\alpha, \beta}I$. 
An inverse formula to that of the Weyl quantization formula was found in 1932 
by Eugene Wigner \cite{Wi32} and maps an operator into what mathematicians 
call its symbol by a kind of trace formula: $\Omega_1$ defines an isomorphism 
of Hilbert spaces between $L^2({\mathbb R}^{2\ell})$ and Hilbert--Schmidt 
operators on $L^2({\mathbb R}^{\ell})$ with inverse given by
\begin{equation} \label{EPW}
u=(2\pi\hbar)^{-\ell}\, {\rm{Tr}}[\Omega_1(u)\exp((\xi.\hat{p}+\eta.\hat{q})/i\hbar)].
\end{equation}

It is important to remember that ``star products" exist as deformations 
(the skew-symmetric part of the leading term being the Poisson bracket $P$) 
of the ordinary product of functions in $N(W)$ for any $W$ \cite{DS01}, 
including when there are no Weyl or Wigner maps and no obvious Hilbert space 
treatment of quantization. They can also be defined for algebraic varieties, 
``manifolds with singularities", and (with some care) infinite-dimensional manifolds. 

We refer, e.g., to \cite{BFFLS,DS01,St12} and especially references therein for 
more developments on deformation quantization and its many avatars. These 
include the notion of covariance of star products and the ``star representations" 
(without operators) it permits. They include also quantum groups, which
appeared in Leningrad around 1980 for entirely different reasons \cite{FRT} but, 
especially after the seminal works of Drinfel'd \cite{Dr86} (who coined the 
name) and of Jimbo \cite{Ji85} (for quantized enveloping algebras) in the 
early 1980s, can be viewed as deformations of (topological) Hopf algebras 
(see, e.g., \cite{BFGP,BGGS,St05}).

In this connection it is worth remembering a prophetic general statement by 
Dirac \cite{Di49}, which applies to many situations in physics: \\
\textsl{Two points of view may be mathematically equivalent, and you may think 
for that reason if you understand one of them you need not bother about the 
other and can neglect it. But it may be that one point of view may suggest a 
future development which another point does not suggest, and although in their 
present state the two points of view are equivalent they may lead to different 
possibilities for the future. Therefore, I think that we cannot afford to 
neglect any possible point of view for looking at Quantum Mechanics and 
in particular its relation to Classical Mechanics.}\\ 
What Dirac had then in mind is certainly the quantization of constrained systems 
which he developed shortly afterward and by now can be viewed as a special case
of deformation quantization. But the principle applies to many contexts and 
is even a most fruitful strategy to extend a framework beyond its initial 
context. A wonderful example is given by noncommutative geometry \cite{Co94}, 
now a frontier domain of mathematics with a wide variety 
of developments ranging from number theory to various areas of physics.        

In order to show that important and concrete problems in physics can be 
treated in an \textsl{autonomous} manner using deformation quantization,
without the need to introduce a Hilbert space (which for most physicists 
is still considered as a requirement of quantum theories) we treated in
\cite{BFFLS} a number of important problems, first and foremost the 
harmonic oscillator (the basic paradigm in many approaches), but also 
angular momentum, the hydrogen atom, and in general the definition of
spectrum inside deformation quantization, without needing a Hilbert space. 
Not so many further applications have been developed since but the approach 
should eventually prove fruitful (even necessary) in many domains in which 
quantum phenomena play a role (including quantum computing). 
True, in many concrete examples we need (at least implicitely) 
``auxiliary conditions" to limit the possibly excessive freedom 
coming, in particular for spectra, from the absence of that Procrustean bed, 
the Hilbert space. But there also, too much freedom is better than not enough.

\subsection{Further important deformations (and contractions)}

\subsubsection{An instance of multiparameter deformation quantization}
A natural question is whether ``the buck stops there" i.e., whether, like for
Gerstenhaber deformations of simple Lie groups or algebras, the structure
obtained is rigid, or whether some further deformations are possible. An
answer to that question, looking for further deformations of $N(W)$ with 
another parameter $\beta$ (in addition to $\nu = i\frac{\hbar}{2}$), was
given in \cite{BFLS} and applied to statistical mechanics and the so-called
KMS states (with parameter $\beta = 1/kT$, $T$ denoting the absolute temperature). 
It turns out that there is some intertwining which is not an equivalence of 
deformations: As a $\nu$-deformation, the two-parameter ``star product" is 
driven by a ``conformal Poisson bracket" with conformal factor of the form 
$\exp (-\frac{1}{2}\beta H)$ for some Hamiltonian $H$.  

\subsubsection{Brief survey of a few aspects of quantum groups} 
\smallskip

The literature on quantum groups (and Hopf algebras) is so vast, diversified 
(and growing) that we shall refer the interested reader to his choice among
the textbooks and papers dealing with the many aspects of that notion, often 
quite algebraic. A two-pages primer can be found in \cite{Md06}.

Roughly speaking quantum groups can often be considered \cite{Dr86} 
as deformations (in the sense of Definition \ref{DrGdef}) 
of an algebra of functions on a Poisson-Lie group (a Lie group $G$ equipped 
with a Poisson bracket compatible with the group multiplication, e.g., 
a semi-simple Lie group), or the ``dual aspect" \cite{Dr86,Ji85} of a 
deformation of (some closure of) its enveloping algebra 
$\mathcal{U}(\mathfrak{g})$ equipped with its natural Hopf algebra structure, 
which is how the whole thing started in Leningrad around 1980. The first 
example was $\mathcal{U}_t(\mathfrak{sl}(2))$, an algebra with generators 
$e,f,h$ as for $\mathfrak{sl}(2)$ but with ``deformed" commutation relations
that can be written somewhat formally (the deformation parameter $t=0$ for 
$\mathfrak{sl}(2)$):
\begin{equation} \label{jqsl2}   
[h,e] = 2e, \ [h,f] = -2f, \ [e,f] = \sinh (th)/\sinh t 
\end{equation}
or more traditionally, as an algebra with generators $E,F,K,K^{-1}$ (one often
writes $K^{\pm 1} = q^{\pm H}$)
and relations
\begin{align}
KK^{-1} = K^{-1}K = 1, \  KEK^{-1} &= q^2E, \  KFK^{-1} =q^{-2}F, \label{1mqsl2} \\
EF - FE &= \frac{K - K^{-1}}{q - q^{-1}},
 \label{2mqsl2} \end{align}
For simplicity we shall not write here the expressions for the coproduct, 
counit and antipode, needed to show the Hopf algebra structure. For higher 
rank simple Lie algebras one has in addition trilinear relations (the deformed 
Serre relations), which complicate matters. All these can be found in the 
literature.    

Note that the algebraic dual of a Hopf algebra is a also Hopf algebra
only when these are finite-dimensional vector spaces, which is quite
restrictive a requirement. In particular the Hopf algebras considered
in quantum groups (except at root of unity), e.g., those of differentiable
functions over a Poisson-Lie group, are (finitely generated) infinite 
dimensional vector spaces; but one can \cite{BFGP,BGGS} define on these spaces 
natural topologies (e.g., Fr\'echet nuclear) which in particular express the 
duality between them and ``quantized 
enveloping algebras". Remember that for any connected Lie group $G$ with 
Lie algebra $\mathfrak{g}$ the elements of the enveloping algebra 
$\mathcal{U}(\mathfrak{g})$ can be considered as differential operators over
$G$, i.e., as distributions with support at any point in $G$ (e.g., the identity 
$e \in G$), which lie in the topological dual of the space of differentiable 
functions with compact support. That exhibits a ``hidden group structure" 
\cite{BFGP} in Drinfeld's quantum groups, which \cite{Dr86} are not groups
(and not always quantum ...) 

\subsubsection{About quantum groups at root of 1} 
As was noticed around 1990, in particular by Lusztig \cite{Lu90a,Lu90g}
(see also, e.g., \cite{DCK,RT91}) the situation changes drastically when the 
deformation parameter is a root of unity. Then the Hopf algebras 
$\mathcal{U}_q(\mathfrak{g})$ become finite dimensional. 

The case of $\mathcal{U}_q(\mathfrak{sl}(2))$ is well understood. For $q$ a 
$2p$-th root of unity, the finite-dimensionality of the algebra comes from
the fact that, in addition to (\ref{1mqsl2}) and (\ref{2mqsl2}), one has 
the relations
\begin{equation} \label{mqsl2f}
K^{2p} = 1, \ E^p = 0, \  F^p = 0.
\end{equation}
In particular (see, e.g., \cite{JX97}) all finite-dimensional indecomposable 
representations have been determined, as well as the indecomposable 
decomposition of tensor products \cite{KSa11}. The higher rank case is 
still largely virgin mathematical territory. It seems that one either needs 
a new approach, or to restrict oneself to particular cases, or both.   
 
\subsubsection{Multiparameter quantum groups}
More generally it is natural to try and look from the start at deformations 
with several scalar parameters. That question seems to have been tackled for 
the first time, in the context of quantum groups, by Manin et~al. 
\cite{DMMZ,Ma91}, who called ``nonstandard" these multiparameter deformations, 
and Reshetikhin \cite{Re90}, and later by Fr\o nsdal \cite{FG95,Fr97}. But
the notion does not seem to have drawn the attention it deserves, 
certainly not much in comparison with the many works on more traditional 
aspects of quantum groups. And more sophisticated questions such as what
happens (outside the generic case), e.g., ``at roots of unity" (whether the 
same root for all parameters or not) do not seem to have ever been 
considered for multiparameter quantum groups. 

\subsubsection{Nonscalar deformation ``parameter"}
Other deformations, more general than those of Gerstenhaber type, were
considered by Pinczon \cite{Pi97} and his student Nadaud \cite{NaDef1,NaDef2}, 
in which the ``parameter" acts on the algebra (on the left, on the right, or 
both) instead of being a scalar. For instance one can have \cite{Pi97}, 
for $\tilde{a}=\sum_n a_n\lambda^n$, $a_n \in A$, a left multiplication by 
$\lambda$ of the form $\lambda \cdot \tilde{a}=\sum_n \sigma(a_n)\lambda^{n+1}$ 
where $\sigma$ is an endomorphism of $A$. A similar deformation theory can be 
done in this case, with appropriate cohomologies, which gives new and 
interesting results. 

In particular \cite{Pi97}, while the Weyl algebra $W_1$ (generated by the 
Heisenberg Lie algebra ${\mathfrak{h}}_1$) is known to be Gerstenhaber-rigid, 
it can be nontrivially deformed in such a 
\textsl{supersymmetric deformation theory} to the supersymmetry
enveloping algebra ${\mathcal{U}}({\mathfrak{osp}}(1|2))$.  
Shortly thereafter \cite{NaDef2}, on the polynomial algebra ${\mathbb C}[x,y]$ 
in 2 variables, Moyal-like products of a new type were discovered; a more 
general situation was studied, where the relevant Hochschild cohomology
is still valued in the algebra but with ``twists" on both sides for the 
action of the deformation parameter on the algebra. 

\subsubsection{Contractions}
Curiously, it is the (less precisely defined) inverse notion of
\textsl{contraction} of symmetries that was first introduced in mathematical 
physics \cite{Se51,IW53}. Contractions, ``limits of Lie algebras" as they were 
called in the first examples, can be viewed as an inverse of deformations 
--~but not necessarily of Gerstenhaber-type deformations. We shall not expand 
on that ``inverse" notion (see \cite{WW00} for a more elaborate study) but
give its flavor since it makes it easier to grasp the deformations of 
symmetries which are important in our presentation. A (finite dimensional) 
Lie algebra $\mathfrak{g}$ can be described in a given basis $L_i$ 
($i=1,\ldots,n$) by its structure constants $C_{i,j}^k$. The equations 
governing the skew-symmetry of the Lie bracket and the Jacobi identity 
ensure that the set of all structure constants lies on an algebraic variety in 
that $n^3$ dimensional space \cite{Ln67}. A contraction is obtained, e.g., when 
one makes a simple basis change of the form $L_i'=\varepsilon L_i$ on 
\textsl{some} of the basis elements, and lets $\varepsilon \to 0$. Take for 
example $n=3$ and restrict to the 3-dimensional subspace of the algebraic 
variety of 3-dimensional Lie algebras with commutation relations 
$[L_1,L_2]=c_3L_3$ and cyclic permutations.
The semi-simple algebras $\mathfrak{so}(3)$ and $\mathfrak{so}(2,1)$ are 
obtained in the open set $c_1c_2c_3\neq 0$. A contraction gives the Euclidean 
algebras, where one $c_i$ is $0$. The ``coordinate axes" (two of the $c_i$'s 
are $0$) give the Heisenberg algebra $\mathfrak{h}_1$ and the origin is the 
Abelian Lie algebra. That is of course a partial picture (e.g., solvable 
algebras are missing) but it is characteristic. 

The above mentioned passage from the Poincar\'e Lie algebra to the Galilean 
is a higher dimensional version of such contractions of Lie algebras 
(multiply the ``Lorentz boosts" generators $M_{0j}$ by $\varepsilon$).
A similar ``trick" on the AdS$_4$ Lie algebra $\mathfrak{so}(3,2)$ gives 
the Poincar\'e Lie algebra. 
A traditional basis for the Poincar\'e Lie algebra is $M_{\mu \nu}$ for the 
Lorentz Lie algebra $\mathfrak{so}(3,1)$ and $P_\mu$ for the space-time 
translations (momentum generators), with $\mu,\nu \in \{0,1,2,3\}$. 
The commutation relations for the conformal Lie algebra $\mathfrak{so}(4,2)$ 
in the basis $M_{\mu \nu}$ with $\mu,\nu \in \{0,1,2,3,5,6\}$ can be written 
(see, e.g., \cite{AFFS})
$[M_{\mu \nu}, M_{\mu' \nu'}] = \eta_{\nu \mu'}M_{\mu \nu'}+ 
\eta_{\mu \nu'}M_{\nu \mu'} - \eta_{\mu \mu'}M_{\nu \nu'} - 
\eta_{\nu \nu'}M_{\mu \mu'}$ with diagonal metric tensor $\eta_{\mu\mu}$ 
(equal, e.g., to +1 for $\mu=0,5$ and -1 for $\mu=1,2,3,6$). One can identify
a Poincar\'e subalgebra by setting, e.g., $P_\mu = M_{\mu 5}+M_{\mu 6}$.
If one omits $\mu = 6$ one obtains the AdS$_4$ Lie algebra $\mathfrak{so}(3,2)$ 
where the role of the Poincar\'e space translations is taken over by $M_{j 5}$ 
($j \in \{1,2,3\}$) and that of the time translations by $M_{05}$.  
How to realize a contraction from AdS$_4$ to Poincar\'e for the massless 
representations used in Section 5 is described in \cite{AFFS}. 
Since it may be easier for physicists to grasp the notion of 
contraction, we mentioned it here and explicited the examples we use in 
familiar notations (possibly old fashioned, but with bases on $\mathbb{R}$).

\section{On the connection between internal and external symmetries}
We shall not extend our desire to be as self contained as possible to 
describing in full detail the Poincar\'e group and its UIRs (unitary 
irreducible representations), known since on the instigation of Dirac 
(his ``famous brother-in-law" as he liked to call him) Wigner \cite{Wi39} 
published the paper that started the study of UIRs of non compact Lie groups. 
Those associated with free particles are usually denoted by $D(m,s)$, 
where $m\geq 0$ is the mass and $s$ the spin of the particle (for $m>0$) 
or its helicity (for $m=0$), associated with the ``squared mass Casimir operator" 
(in the center of the Poincar\'e enveloping algebra) $P_\mu P^\mu$ and
with the inducing representation of the ``little group" ($SO(3)$ and 
$SO(2)\cdot \mathbb{R}^2$, resp.). [The letter $D$, coming from 
the German ``Darstellung", is often used to denote representations.]
       
In the early 1960s a natural question appeared: Is there any connection 
between the ``internal symmetry" used in the classification of interacting 
elementary particles (tentatively $SU(3)$ at the time), and the Poincar\'e 
symmetry whose UIRs are associated with free particles? 
The question is not innocent since 3 octets of different spins were associated 
with the same representation of $SU(3)$ (the 8-dimensional adjoint 
representation). And the various families within an octet (the same applies
to the decuplet) exhibit a mass spectrum. If there is a connection, one ought 
to describe a mechanism permitting all that. Of course an important issue is 
how to mathematically formulate the question.    

\subsection{No-go theorems, objections, counter-examples and generalizations}
\subsubsection{A Lie algebra no-go and counterexamples}                    
In the ``particle spectro\-scopy" spirit of the time, it was natural to
look for a Lie algebra containing both symmetries (internal and external).
In 1965, a year after quarks and color were proposed, appeared a ``no-go 
theorem" as physicists like to call such results, due to L. O'Raifeartaigh 
\cite{O'R65}. It boiled down to the fact that, since the momentum generators 
$P_\mu$ are nilpotent in the Poincar\'e Lie algebra, they are nilpotent in
any simple Lie algebra containing it, which forbids a discrete mass spectrum. 
Hence in order to have a mass spectrum the connection must be a direct sum. 
Almost everybody was happy, except that two trouble makers in France said:
``It ain't necessarily so". Thanks to Isidor Rabi (then president of the
APS, who remembered Moshe Flato as a most brilliant student who often asked 
difficult questions during the course he gave for a quarter at the Hebrew
University, invited by Giulio Racah) our objection was published shortly
afterward \cite{FS65} in the provocative form desired by Moshe. It was 
followed by counterexamples \cite{FS66,FS69}. The problem with the ``proof" 
in \cite{O'R65} is that it implicitly assumed the existence of a common 
invariant domain of differentiable vectors for the whole Lie algebra, 
something which Wigner was careful to state as an assumption in \cite{Wi39} 
and was proved later for Banach Lie group representations by (in Wigner's own 
words) ``a~Swedish gentleman" \cite{Go47}. Eventually the statement of 
\cite{O'R65} could be proved within the context of UIRs of finite dimensional 
Lie groups \cite{Jo66} and was further refined by several authors (especially 
L.~O'Raifeartaigh). However we showed in \cite{FS69} that a mass spectrum is 
possible when assuming only the Poincar\'e part to be integrable to a UIR, 
and there is no a priori reason why the additional observables should close to 
a finite dimensional group UIR. We gave also counterexamples \cite{FS69i} with 
a natural infinite-dimensional group and even showed \cite{FS67} that it is 
possible to obtain any desired mass spectrum in the framework of 
finitely-generated infinite-dimensional associative algebras and unitary groups. 

Like with many physical ``theorems", a main issue is to decide what assumptions 
and what heuristic developments can be considered as ``natural". While some 
flexibility can be accepted in proving positive results, no-go theorems should 
be taken with many grains of salt.      

\subsubsection{Subsequent developments and relative importance of the question}
As we indicated in \cite{FS69}, such considerations apply also to a more 
sophisticated no-go theorem \cite{CM67}, formulated in the context of 
symmetries of the $S$-matrix, which could be applied also to infinite-dimensional
groups. This very nice piece of work is still considered by most physicists 
(especially those who learned it at university) as definitely proving the 
direct sum connection (under hypotheses easily forgotten, some of which are 
even hidden in the apparently natural notations).  

In retrospect one can say that a main impact of the latter result came 
through an attempt to get a supersymmetric extension \cite{HLS75}, which showed 
one might get around the no-go in the supersymmetry context. That gave a big 
push to the latter. Incidentally it is generally considered that the 
``super-Poincar\'e group" of Wess and Zumino \cite{WZ74} is practically the 
first instance of supersymmetry. That is not quite correct. In particular 
already in 1967 (in CRAS) we introduced what we called ``a Poincar\'e-like 
group", semi-direct product of the Lorentz group and $\mathbb{R}^8$ consisting 
of both vector and spinor translations, but (for fear of Pauli) we did not 
dare introduce anticommutators for spinorial translations together with 
commutators for space-time translations, so we remained in the Lie algebra 
framework. However one can find in \cite{FH70} a physical application of that 
group in which the spinorial translations are multiplied by an operator~$F$ 
anticommuting with itself. That was in effect the first realization of the 
super-Poincar\'e group. Both Wess and Zumino told me some years ago that they
were unaware of the fact and it seems that (except for Fr\o nsdal) not many 
noticed it either.  

\section{Singleton physics}
\subsection{Singletons as ``square roots" of massless particles}
The contraction of AdS to Poincar\'e (in both structure and representations)
is (cf. \cite{AFFS}) one of the justifications for calling ``massless" 
some minimal weight UIRs of $Sp(\mathbb{R}^4)$ (the double covering of 
$SO(3,2)$). These are denoted by $D(E_0,s)$, the parameters being the lowest 
values of the energy and spin (resp.) for the compact subgroup 
$SO(2)\times SO(3)$. These irreducible representations are unitary provided 
$E_0 \geq s+1$ for $s\geq 1$ and $E_0 \geq s+\frac{1}{2}$ for $s=0$ 
and $s=\frac{1}{2}$. The massless representations of ${SO}(3,2)$ are
thus defined (for $s\geq \frac{1}{2}$) as $D(s+1,s)$  and (for helicity zero) 
$D(1,0)\oplus D(2,0)$. At the limit of unitarity (when going down in the
values of $E_0$ for fixed $s$) the Harish Chandra module $D(E_0,s)$ becomes 
indecomposable and the physical UIR appears as a quotient, a hall-mark of 
gauge theories. For $s\geq 1$ we get in the limit an indecomposable 
representation $D(s+1,s)\leadsto D(s+2,s-1)$, where $\leadsto$ (``leaking into")
is a shorthand notation \cite{FF88} for what mathematicians would write as 
a short exact sequence of (infinite-dimensional) modules.

A complete classification of the UIRs of the (covering of) $SO(p,2)$ can be 
found, e.g., in \cite{An81}. For $p=3$ the classification had been completed 
when Dirac \cite{Di63} introduced the most degenerate ``singleton" 
representations. The latter are irreducible and massless on a subgroup, the 
Poincar\'e subgroup of a $2+1$ dimensional space-time, of which AdS is the 
conformal group. That is why (on the pattern of Dirac's ``bra" and ``ket") 
we call these representations $Di=D(1,\frac{1}{2})$ and $Rac=D(\frac{1}{2},0)$ 
for (resp.) the spinorial and scalar representations. 
The singleton representations have a fundamental property:\\
\begin{equation}\label{di+rac}
(Di\oplus Rac)\otimes(Di\oplus Rac)=(D(1,0)\oplus D(2,0))\oplus 
2 \bigoplus_{s=\frac{1}{2}}^\infty D(s+1,s).
\end{equation}
The representations appearing in the decomposition are what we call 
massless representations of the AdS group, for a variety of 
good reasons \cite{AFFS}. For $s=0$ a split occurs because time is compact 
in AdS and one does not distinguish between positive and negative
helicity, which is also the reason for the factor 2 in front of the sum. 
An extension to the conformal group $SO(4,2)$, which is operatorially
unique for massless representations of the Poincar\'e group and (once a 
helicity sign is chosen) for those of $SO(3,2)$, or a contraction of the latter
to the Poincar\'e group, restores the distinction between both helicity signs 
and provide other good reasons for calling massless representations of AdS 
those in the right hand side of (\ref{di+rac}).   

Thus, in contradistinction with flat space, in AdS$_4$, massless states
are ``composed'' of two singletons. The flat space limit of a singleton
is a vacuum and, even in AdS$_4$, the singletons are very poor in states: 
their $(E,j)$ diagram has a single trajectory (hence the name given to them 
by Dirac), and is not a lattice like, e.g., for massless particles in AdS. 
In normal units a singleton with angular momentum $j$ has energy
$E=(j+\frac{1}{2})\rho$, where $\rho$ is the curvature of the
AdS$_4$ universe. This means that only a laboratory of cosmic
dimensions can detect a $j$ large enough for $E$ to be measurable:
one can say that the singletons are ``naturally confined". 

Like the AdS$_n$/CFT$_{n-1}$ correspondence, the symmetry part of which states
essentially that $SO(n-1,2)$ is the conformal group for $n-1$-dimensional 
space-time, singletons exist in any space-time of dimension $n \geq 3$ 
\cite{AL00}, $n=4$ being somewhat special. For $n=3$ the analogue of 
(\ref{di+rac}) writes $(HO)\otimes (HO) = Di \oplus Rac$ where $(HO)$ denotes 
the harmonic oscillator representation of the metaplectic group (double 
covering of $SL(2,\mathbb{R})$, itself a double covering of $SO(2,1)$) which 
is the sum of the discrete series representation $D(\frac{3}{4})$ and of the 
complementary series representation $D(\frac{1}{4})$. One thus has a kind 
of ``dimensional reduction" by which ultimately massless particles can be 
considered as arising from the interaction of harmonic oscillators. 
I leave it to the reader to derive consequences from that fact, and maybe study 
connections with the challenging suggestions of Gerard 't Hooft (see, e.g., 
\cite{tH13a,tH13b} and references therein) dealing with ``quantum determinism"
and based in particular on cellular automata.  

\textbf{Remark 5.1.1. Phase space realization.} Quadratic polynomials 
in $2\ell$ real variables $p_\alpha$ and 
$q_\beta$ ($\alpha,\beta \in \{1, \ldots, \ell\}$), satisfying the Heisenberg 
canonical commutation relations (CCR) 
$[p_\alpha, q_\beta] = \delta_{\alpha\beta} I$, 
generate a realization of the symplectic Lie algebra 
$\mathfrak{sp}(\mathbb{R}^{2\ell})$. Together with the linear polynomials 
(under anticommutators) they close to an irreducible realization of 
$\mathfrak{osp}(1|2\ell)$, the corresponding superalgebra 
($\mathbb{Z}_{(2)}$-graded). [$Di\oplus Rac$ and $(HO)$ are special cases of
the phenomenon for $\ell = 2,1$ (resp.).] That fact allowed us in \cite{BFFLS} 
to write a power series expansion in $t$ of what we call the ``star exponential" 
$\mathrm{Exp}*(tH/i\hbar)$ (corresponding in Weyl quantization to the unitary 
evolution operator) of the harmonic oscillator Hamiltonian 
$H = \frac{1}{2}(p^2 + q^2)$, while a theorem of Harish Chandra states that 
the character of a UIR (which in our formalism is the integral over phase space 
of the star-exponential) always has a singularity at the origin: 
the singularities for the two components in $(HO)$ cancel at the origin, a 
true miracle which puzzles many specialists of Lie group representation theory!        

\subsection{AdS$_4$ dynamics}
Until now we were concerned mainly with what can be called the 
``kinematical aspect" of the question, i.e., symmetries. However at some point
one cannot avoid looking at the dynamics involved. In particular covariant
field equations and Lagrangians will have to be studied. And indeed many 
papers were written, especially in the 1980s and 1990s by Flato, Fr\o nsdal 
and coworkers, developing various aspects of singleton physics. These include
BRST symmetry, conformal aspects and related indecomposable representations 
(in particular of the Gupta--Bleuler type), etc. Our purpose here is to build 
on these and on the ``deformation philosophy" and not to give an extensive 
account of all these works, references to many of which can be found in Flato's
last paper \cite{FFS99}. In the next two subsections we shall give a brief 
account of the two papers that are the most important from the point of view 
developed here, composite QED \cite{FF88} and what can be called an 
electroweak model extended to 3 generations of leptons \cite{Fr00}. 
 
\subsubsection{The Flato--Fr\o nsdal ``singletonic QED"}
Dynamics require in particular the consideration of field equations,
initially at the first quantized level, in particular the analogue of the
Klein--Gordon equation in AdS$_4$ for the $Rac$. There, as can be expected
of massless (in 1+2 space) representations, gauges appear, and the physical
states of the singletons are determined by the value of their fields on
the cone at infinity of AdS$_4$ (see below; we have here a phenomenon of
holography \cite{tH93}, in this case an AdS$_4$/CFT$_3$ correspondence).

We thus have to deal with indecomposable representations, triple extensions
of UIR, as in the Gupta--Bleuler (GB) theory, and their tensor products.
[It is also desirable to take into account conformal covariance at
these GB-triplets level, which in addition permits distinguishing between
positive and negative helicities (in AdS$_4$, the time variable being
compact, the massless representations of $SO(2,3)$ of helicity $s > 0$
contract (resp. extend in a unique way) to massless representations of
helicity $\pm s$ of the Poincar\'e (resp. conformal) group.]
The situation gets therefore much more involved, quite different from
the flat space limit, which makes the theory even more interesting.

In order to test the procedure it is necessary to make sure that it is
compatible with conventional Quantum Electrodynamics (QED), the best
understood quantum field theory, at least at the physical level of rigor.

One is therefore led to see whether QED is compatible with a massless photon
composed of two scalar singletons. For reasons explained, e.g., in \cite{FFS99}
and references quoted therein, we consider for the $Rac$, the dipole equation
$( \square\ -\frac {5}{4}\rho)\ ^{2}\phi\ = 0$ with the boundary
conditions $r^{1/2}\phi\ <\infty$ as $r\rightarrow\infty$, which
carries the indecomposable representation                            
$D(\frac {1}{2}, 0) \ \leadsto \ D(\frac {5}{2},0)$.
A remarkable fact is that this theory is a  \textsl{topological field theory};
that is \cite{FF81}, the physical solutions manifest themselves only by
their boundary values at $r\rightarrow\infty$: $\lim r^{1/2}\phi$
defines a field on the 3-dimensional boundary at infinity.  There, on the
boundary, gauge invariant interactions are possible and make a 3-dimensional
conformal field theory (CFT).

However, if massless fields (in four dimensions) are singleton composites,
then singletons must come to life as 4-dimensional objects, and this
requires the introduction of unconventional statistics (neither Bose--Einstein
nor Fermi--Dirac).  The requirement that the bilinears have the properties of
ordinary (massless) bosons or fermions tells us that the statistics of 
singletons must be of another sort. The basic idea is \cite{FF88} that we can 
decompose the $Rac$ field operator as                                                      
$\phi(x)=\sum_{-\infty}^\infty \phi^j(x)a_j$ in terms of positive
energy creation operators $a^{*j}=a_{-j}$ and annihilation
operators $a_j$ (with $j>0$) without so far making any assumptions
about their commutation relations. The choice of commutation relations comes
later, when requiring that photons, considered as 2-$Rac$ fields, be
Bose--Einstein quanta, i.e., their creation and annihilation operators
satisfy the usual canonical commutation relations (CCR). The singletons
are then subject to unconventional statistics (which is perfectly
admissible since they are naturally confined), the total algebra being
an interesting infinite-dimensional Lie algebra of a new type,
a kind of ``square root" of the CCR. An appropriate Fock space can then
be built. Based on these principles, a (conformally covariant) composite QED
theory was constructed \cite{FF88}, with all the good features
of the usual theory---however years after QED was developed by
Schwinger, Feynman, Tomonaga and Dyson.

\textbf{Remark 5.2.1.1. Classical Electrodynamics} as a covariant nonlinear 
PDE approach to coupled Maxwell--Dirac equations.
Only relatively recently was classical electrodynamics (on 4 dimensional
flat space-time) rigorously understood. By this we mean the proof of 
asymptotic completeness and global existence for the coupled Maxwell--Dirac 
equations, and a study of the infrared problem. That was done \cite{FST97} 
with the third aspect of our trilogy (complementing deformation quantization 
and singleton physics), based on a theory of nonlinear group representations, 
plus a lot of hard analysis using spaces of initial data suggested by the 
linear group representations. The deformation quantization of that 
classical electrodynamics (e.g., on an infinite dimensional phase space of 
initial conditions) remains to be done. 

\subsubsection{Composite leptons, Fr\o nsdal's extended electroweak model}
A natural step, after QED, is to introduce compositeness in electroweak
theory. Along the lines described above, that would require finding
a kind of ``square root of an infinite-dimensional superalgebra,"
with both CAR (canonical anticommutation relations) and CCR included:
The creation and annihilation operators for the naturally confined $Di$
or $Rac$ need not satisfy CAR or CCR; they can be subject to unusual
statistics, provided that the two-singleton states satisfy Fermi--Dirac
or Bose--Einstein statistics depending on their nature.
We would then have a (possibly $\mathbb{Z}$-)graded algebra where
only the two-singleton states creation and annihilation operators satisfy
CCR or CAR. That has yet to be done. Some steps in that direction have been
initiated but the mathematical problems are formidable, even more so since
now the three generations of leptons have to be considered.

But here a more pragmatic approach can be envisaged \cite{Fr00}, triggered
by experimental data showing oscillations between various generations of 
neutrinos. The latter can thus no more be considered as massless.
This is not as surprising as it seems from the AdS point of view, because
one of the attributes of masslessness is the presence of gauges. These
are group theoretically associated with the limit of unitarity in the
representations diagram, and the neutrino is above that limit in AdS: the
$Di$ is at the limit. Thus, all nine leptons can be treated on an 
equal footing.

It is then natural \cite{Fr00} to arrange them in a square table ($L^A_\beta$), 
the rows being the 3 generations of leptons, each of which carry the Glashow 
representation of the `weak group' $S_W = SU(2)\otimes U(1)$ and to consider 
the 9 leptons $L^A_\beta$ ($\nu_e, e_L, e_R$ and similarly for the two other 
generations $\mu$ and $\tau$) as composites, $L^A_\beta = R^A D_\beta$ 
($A = N,L,R; \beta = \epsilon, \mu, \tau$). We assume that the $R^A$s are 
$Rac$s and carry the Glashow representation of $S_W$, while the 
$D_\beta$s are $Di$s, insensitive to $S_W$ but transforming as a Glashow 
triplet under a `flavor group' $S_F$ isomorphic to $S_W$. To be more 
economical we also assume that the two $U(1)$s are identified, the same 
hypercharge group acting on both $Di$s and $Rac$s. As explained in \cite{Fr00},
the leptons are initially massless (as $Di$-$Rac$ composites) and  
massified by (in effect, five) Higgs fields $K^{\alpha\beta}_{AB}$ that 
(like in the electroweak model) have a Yukawa coupling to the leptons. 
The model predicts, in parallel to the $W^\pm$ and $Z$ bosons, two new bosons 
$C^\pm$ and $C^3$ (hard to detect due to the large mass differences between 
the 3 generations of leptons) and explains the neutrino masses. It is 
qualitatively promising but the presence of too many free parameters limits 
its quantitative predictive power. 

One could be tempted to add to the picture a deformation induced by 
the strong force and 18 quarks, which (with the 9 leptons) could be written
in a cube and also considered composite (of maybe three constituents when
the strong force is introduced). That might make this ``composite Standard Model"
more predictive. But introducing the hadrons brings in a significant 
quantitative change that should require a qualitative change, e.g., 
some further deformation of the AdS symmetry.

\section{Hadrons and quantized Anti de Sitter}
\subsection{A beginning of a new picture}
Instead of a ``totalitarian" approach aiming towards a ``theory of everything,"
at least inasmuch as elementary particles are concerned, we shall adopt an
approach which is both more pragmatic and based on fundamental principles.
Since symmetries were the starting point from which what is now the Standard
Model emerged, and since free particles are governed by UIRs of the Poincar\'e
group (the symmetry of special relativity), we shall start from the latter
and proceed using our ``deformation philosophy" as a guideline. 

\subsubsection{Photons, Leptons and Hadrons.} 
As we have seen, assuming that in the ``microworld", i.e., at some scale 
(to be made more precise eventually) the universe is endowed with a small
negative curvature, the Poincar\'e group is deformed to AdS and the massless
photon states can be dynamically (in a manner compatible with QED) considered 
as a 2-$Rac$ state \cite{FF88}. Then, extending the electroweak theory to the 
empirically discovered 3 generations, the 9 leptons can be considered, using
the AdS deformation of the Poincar\'e group, as initially massless $Di$-$Rac$ 
states, massified by 5 Higgs bosons \cite{Fr00}.

The ``tough cookie" is then how to explain hadrons and strong interactions.
The deformation philosophy suggests to try and deform AdS, which is not
possible as a group but can be done as a Hopf algebra, to a ``quantum group",
qAdS. In the ``generic case" the obtained representation theory will not be 
very different from the AdS case. That is essentially due to the ``Drinfeld 
twist" which intertwines between AdS and qAdS, even if that is not an 
equivalence of deformations (it is a kind of ``outer automorphism"). 
But at root of unity the situation becomes drastically different: the 
Hopf algebra becomes finite-dimensional (which is not the case of the generic
qAdS) and there are only a finite number of irreducible representations. 

\subsubsection{Remarks on the mathematical context.} 
The fact that only a few irreducible representations may be relevant is both 
an encouraging feature and a restrictive one, even more so since for quantum
groups at root of 1, the theory of tensor products of such representations 
(needed in order to consider these as describing interacting particles), which
is ``nice" in the generic case, is not straightforward. 
In particular the tensor products are usually indecomposable, extensions of 
direct sums of irreducibles defined by some cocycles (which could however 
be related to the ``width" of the observed resonances). The phenomenon appears 
already in rank one (quantized $\mathfrak{sl}(2)$ at root of unity) \cite{KSa11}, 
where the category of representations is not braided and the tensor products 
$R\otimes S$ and $S\otimes R$ of two representations can sometimes be different.                                                    
The general theory for higher ranks seems hopeless. But for
physical applications we do not need a general theory, possibly on the 
contrary: If only a few representations behave ``nicely", and it turns out 
that Nature selects these, so much the better. [Even for the Poincar\'e group
only half of the UIRs are physically relevant, those of positive mass, and
of zero mass and discrete helicity.] The case of rank 2 (in particular qAdS) 
seems more within reach, though the mathematics is highly non trivial. 
Some works are in progress in that direction, in particular by Jun Murakami 
who recently \cite{CM13} studied quantum $6j$-symbols for $SL(2,\mathbb{C})$.          
But a lot remains to be done in order to clarify the mathematical background.
   
\subsection{Quantized AdS (in particular at some root of 1) and generalizations} 
\subsubsection{Some ideas, problems and results around qAdS representations}
Over 20 years ago appeared a concise and interesting paper \cite{FHT93}, 
written, without sacrificing rigor, using a language (e.g., Bose creation and 
annihilation operators, supersymmetry, Fock space) that can appeal to physicists
(but maybe less to mathematicians \ldots). In that paper, on the basis of a 
short panorama of singleton and massless representations of $\mathfrak{so}(3,2)$, 
their supersymmetric extensions and (for the massless) imbedding in the 
conformal Lie algebra $\mathfrak{u}(2,2)$, the authors dealt with 
$q$-deformations of that picture, especially $q$-singletons, $q$-massless 
representations, and the imbedding therein of $q$-deformations of 
$\mathfrak{sl}(2)$. They studied both the case of generic $q$ and the case 
when $q$ is an even root of unity. A main purpose was, in the latter case, to
write explicitly, in the case of $\mathcal{U}_q(\mathfrak{so}(3,2))$, defining 
relations similar to those of (\ref{1mqsl2}) and (\ref{2mqsl2}) (including now 
the ``$q$-Serre relations") and to express in a more physical language the fact 
(discovered a few years before by Lusztig \cite{Lu90g}) that one gets then 
finite-dimensional unitary representations.    

Incidentally the ``unitarization" of irreducible representations can be 
important for possible physical applications. For quantum groups at root of 
unity that has been studied, in particular for AdS, in \cite{FHT93,Sa98} and 
for many series of simple noncompact Lie algebras in \cite{Sa01}, where 
in addition it is shown that the unitary highest weight modules of the 
classical case are recovered in the limit $q \to 1$.  

Guided by the ``deformation philosophy" we are thus led to look at what happens 
when AdS, the deformation of the Poincar\'e group when we assume a (tiny) 
negative curvature in some regions, is further deformed to the 
quantum group qAdS. The idea is that ``internal symmetries" might arise as 
such deformations, which seems especially appealing at root of unity because
we have then finite-dimensional unitary representations.  

Now if we want to try and assign multiplets of particles to (irreducible) 
representations of qAdS at root of 1, a first step is to know what are the 
dimensions of these representations. These dimensions have been found 
(by Jun Murakami, work in progress) for sixth root of 1, to be: 1; 4,5; 
10,14,16; 35,40; 81. We chose here $p=3$ in the $2p^\mathrm{th}$ root of 1, 
physically because there are 3 generations, and mathematically because for a 
variety of reasons one must take $p\geq 3$, so that is the first case.

The first nontrivial representations are, as can be expected in the case of
a Lie algebra of type $B_2 \equiv C_2$, of dimensions 4 and 5 (that was 3 in
the case of $\mathfrak{su}(3)$, hence quarks). Thus if we want to mimic what 
has been done for unitary symmetries, we might have to replace, e.g., the basic 
octet by two ``quartets", unless a doubling of the dimension can be justified 
by some mathematical or physical reasons (maybe looking at a corresponding supersymmetry). 

Note that while all (compact) simple groups of rank 2 were studied in detail 
in \cite{BDFL62} from the point of view of strong interaction symmetries 
and eventually type $A_2$ emerged, and while the (finite-dimensional) 
representation theory of generic quantum groups is similar to the classical 
case, the restricted quantum groups are so different that the $B_2$ type 
cannot be excluded a priori. A notable difference is that the restricted 
quantum groups have only a finite number of finite-dimensional representations,
which might be an advantage. 

But the knowledge of the dimensions is only the beginning of the beginning. 
In particular we need to study the tensor products of the representations 
we want to use, which helps to describe what happens when two particles 
interact strongly and eventually produce other particles. 
In the case of roots of unity these tensor products typically give rise to 
indecomposable representations, essentially extensions of irreducibles 
given by some cocycles (somewhat like in the Gupta-Bleuler formalism for the 
electromagnetic field). The fact might here be related to the widths of the 
resonances produced, but that is so far only a conjecture.  

The general study of such tensor products (beyond the rank 1 case, where 
it is already complicated) is nontrivial mathematically. As a ``warm up 
exercise" for rank 2, it may be worth to start with the $A_2$ case, 
i.e., $\mathcal{U}_q(\mathfrak{sl}(3))$. 

If we want to have for strongly interacting particles a picture similar 
to what has been done so far with unitary symmetries, we could first want to 
assign particle multiplets with some low-dimensional representations of qAdS. 
But since now we have a connection (via deformations) between the free 
(Poincar\'e) and the strongly interacting symmetry (qAdS), which conceptually 
is an advantage, we should imagine a mechanism explaining why to assign some 
spins (traditionally associated with the Poincar\'e group) to such multiplets, 
and how can we have a mass spectrum inside the multiplet. 

Assuming we solve these (hard) ``mathematical and physical homeworks", the 
physical task ahead of us is even more formidable: Re-examine critically 
half a century of particle physics, first from the phenomenological and 
experimental points of view on the basis of the new symmetries. 
``Going back to the drawing board," we should then re-examine the present 
phenomenology in the new framework, including interpretations of raw 
experimental data. These were so far made in the context of the standard 
model and the quarks hypothesis, starting from nucleons and a few other 
particles and explaining inductively the observations in accelerators and 
cosmic rays within that framework. 

Note that, as we have seen in Section 5, a main success of the present 
theory (the electroweak model) is preserved by deforming Poincar\'e to AdS. 
The speculations in this subsection are natural extensions of that in order 
to try and describe strong interactions using our deformation philosophy.   

Inasmuch as we would ``simply" replace the internal symmetries by some qAdS 
at root of 1, we should also, from the theoretical point of view, re-examine 
the various aspects (e.g., QCD) of the dynamics that was built on the possibly 
``clay feet" of simple unitary symmetries. That is a colossal task, 
but not as much as it may seem because we are not starting from scratch. 
A lot of the sophisticated notions introduced and theoretical 
advances made in the past decades might be adapted to our ``deformed" view
of symmetries. That could include many parts of the string framework.  

Even more so since in the same spirit, it is possible that more sophisticated 
(and largely unexplored) mathematics would require less drastic a departure 
from the present puzzle, the pieces of which fit so well (so far). We shall 
explain that in the following.  

\subsubsection{Generalizations: Multiparameter, superizations and affinizations}
An essential part in the representation theory of the traditional internal 
symmetries like $SU(n)$ (we take usually $n=3$) boils down to questions of 
number theory, around the Weyl group ($\mathbb{S}_n$ for $SU(n)$) and the 
center of the group ($\mathbb{Z}_{(n)}$ in that case). One is therefore 
led to study what can be said of quantum deformations of 
$\mathcal{U}(\mathfrak{so}(3,2))$ when we take for deformation parameter 
an element of the group algebra of the center, $\mathbb{C}\mathbb{Z}_{(n)}$. 
That is, if we want to remain in the context already studied (see the next 
section for a more daring suggestion) of multiparameter deformations. 
While the generic case is relatively well understood (cf. Section 3.3.4), 
the case of root of 1 seems not to have been considered. It is not even clear 
whether one could (or should) take the same root of 1 for all the generators 
of the center. The ``warping effect" of the roots of 1, which manifests itself 
already in (\ref{mqsl2f}), could play tricks. The same procedure can be 
applied to quantizing the superalgebra $\mathfrak{osp}(1|4)$ obtained 
\cite{Fr82,FHT93} from the realization of $\mathfrak{so}(3,2))$ as quadratic 
homogeneous polynomials in 4 real variables ($p_1,p_2;q_1,q_2$) by adding 
the linear terms (endowed with anticommutators).

In view of possibly incorporating dynamics into the symmetry picture, and in 
the spirit of the string framework of ``blowing up" points, often a cause of 
singularities, into, e.g., strings, one may then want to consider loop algebras
(maps from a closed string $S^1$ to the symmetry in question, e.g., 
$\mathfrak{so}(3,2)$), and their quantization. In the same vein one may
want to consider a number of infinite-dimensional algebras (Kac--Moody, 
Virasoro, etc.) built on that pattern. That is a very active topic (cf. e.g.,
\cite{HJ12,JR12}) in which many results are available. But the kinds         
of specific examples we would need here have not been much studied and the 
root of unity case even less (mildly speaking). And then one may want to say 
something about the very hard question of maps from something more general 
than $S^1$ (e.g., a $K_3$ surface or a Calabi--Yau complex 3-fold) into some 
groups or algebras, and the possible quantization of such structures: 
these are totally virgin territory.     

All these generalizations, even in the specific cases we would need here for
possible applications, are at least valid mathematical questions.   
  
\subsubsection{``Quantum deformations" (with noncommutative ``parameter")}
The  ``best of both worlds" might however result from a challenging idea.  
Internal symmetries could emerge from deforming those of space-time in a 
more general sense that would include the use made of unitary symmetries 
like $SU(n)$. A fringe benefit might even be to give a conceptually 
beautiful explanation to the fact that we observe 3 generations. 

The idea would be to ``quantize" the Gerstenhaber definition (\ref{DrGdef}) of 
deformations of algebras, not simply by considering a parameter that acts on 
the algebra (as indicated in Section 3.3.5) but by trying to develop a similar
theory with a noncommutative ``parameter". That has not been done and is 
far from obvious. It is not even clear what (if any) cohomology would be needed.

In particular one may think of deformations with a quaternionic deformation
parameter. The field $\mathbb{H}$ of quaternions is the only number field 
extending that of complex numbers $\mathbb{C}$, but it is nonabelian. 
So we could speak of such a theory as a ``quantum deformation" since one often  
calls ``quantum" mathematical notions that extend existing ones by 
``plugging in" noncommutativity. As is well known, elements of $\mathbb{H}$ 
can be written in the form $a+bi+cj+dk$ with $a,b.c.d \in \mathbb{R}$, 
$i^2=j^2=k^2=-1$ and $i,j,k$ anticommuting (like Pauli matrices). 
Interchanging the roles of $i,j,k$ would give a symmetry ($\mathbb{S}_3$ or 
$SO(3)$) which might explain why we have 3 (and only 3) generations. 
And deforming $\mathcal{U}(\mathfrak{so}(3,2))$ (the choice of that real form
could be important) using such a (so far, hypothetical) quantum deformation 
theory, especially in some sense at root of unity, might give rise to internal 
symmetries for which the role of the Weyl group $\mathbb{S}_3$ of $SU(3)$ 
(and possibly of $SU(3)$) would permit to re-derive, this time on a 
fundamental basis, all what has been done with the representations of $SU(3)$. 
We would still have to explain, e.g., spin assignments to multiplets, 
and much more. But now we would have a frame for that, and a subtle nontrivial 
connection between internal and external symmetries could be developed, 
with all its implications (especially concerning the dynamics involved), 
with a relatively modest adaptation of the present empirical models. 

A variant of that would be to develop a theory of deformations parame\-trized 
by the group algebra of $\mathbb{S}_n$. Except for the fact that we would have 
to assume, e.g., $n=3$ and not ``explain" why we have only 3 generations, the
general idea would be similar, so we shall not repeat the above speculations.

More generally, a mathematical study of such ``quantum deformations" should be
of independent interest, even if that uses very abstract tools that may not 
make it directly applicable to the physical problems we started from. 

\subsubsection{Remark: Quantized AdS space and related cosmology}
One should not only try and develop a theory of the fundamental constituents
of the matter we know (even if it constitutes only about 4\% of the universe)
but also explain how that small part is created, and if possible how
comes that we see so little antimatter. To this end also, quantizing AdS 
space-time might help. 

Of course, in line with recent observational cosmology, our universe is 
probably, ``in the large", asymptotically de Sitter, with positive curvature
and invariance group $SO(4,1)$, the other simple group deformation of the
Poincar\'e group which however, unlike AdS, does not give room to a positive 
energy operator. At our scale, for most practical purposes, we can treat 
it as Minkowskian (flat). Focussing ``deeper" we would then discover that
it can be considered as Anti de Sitter. There one can explain photons and 
leptons as composites of singletons that live in AdS space-time. It is thus 
natural to try and quantize that AdS space-time. And in fact, in \cite{BCSV}, 
we showed how to build such ``quantized hyperbolic spheres", i.e., 
noncommutative spectral triples \`a la Connes, but in a Lorentzian context, 
which induce in (an open orbit in) AdS space-time a pseudo-Riemannian 
deformation triple similar (except for the compactness of the resolvent) to the
triples developed for quantized spheres by Connes et al. (see, e.g., \cite{CD03}). 
Such a ``quantized AdS space" has a horizon which permits to consider it 
as a black hole (similar to the BTZ black holes \cite{BHTZ}, which exist 
for all AdS$_n$ when $n\geq3$). [A kind of groupoid structure might be
needed if one wants to treat all 3 regions.] 

For $q$ an even root of unity, since the corresponding quantum AdS group has 
finite dimensional UIRs, such a quantized AdS black hole could be considered
as ``$q$-compact" in a sense to be made precise. As we mention in 
\cite{BCSV,St07}, in some regions of our universe, our Minkowski
space-time could be, at very small distances, both deformed to anti de Sitter 
and quantized, to qAdS. These regions would appear as black holes which
might be found at the edge of our expanding universe, a kind of ``stem cells"
of the initial singularity dispersed at the Big Bang. From these (that is so 
far mere speculation) might emerge matter, possibly first some kind of 
singletons that couple and become massified by interaction with, e.g., dark matter 
and/or dark energy. Such a scheme could be responsible, at very large 
distances, for the observed positive cosmological constant -- and might 
bring us a bit closer to quantizing gravity, the Holy Grail of modern physics,
whether or not that is a relevant question (even if very recent and well publicized 
observations of gravitational waves might indicate that quantizing gravity is needed).

\section{Epilogue and a tentative ``road map"}
After such a long overview involving a fireworks of fundamental mathematical 
and physical notions, many of which need to be developed, a natural question
(which most physicists will probably ask after going through the abstract) is:
why argue with success? 

After all, the Standard Model is considered (so far) as the ultimate 
description of particle physics.    
We were looking for a key to knowledge under a lantern (bei der Laterne), 
found one which turned out to open a door nearby (not the one sought initially 
but never mind) through which a blue angel M led us to a beautiful avenue 
with many ramifications. But what if M was Fata Morgana and that avenue 
eventually turns out to be a dead end? We would then need a powerful 
flashlight (deformations maybe) to find in a dark corner a strange key with
which, with much effort, a hidden door can be opened and lead us to an avenue 
where part of our questions can be answered.   
  
Still, physicists tend to have a rather positivist attitude and, 
in most areas of physics, one takes for granted some experimental 
facts without trying to explain them on the basis of fundamental principles. 
Why would particle physics be different? A first answer is given by Einstein: 
``Curiosity has its own reason for existing." Theoretical particle physics 
is certainly an area in which very fundamental questions can be asked, and 
answered, if needed with the help of sophisticated mathematics to be 
developed. 

So, at the risk of being considered simple minded, I am asking
the question of \textsl{why} the symmetries on which is based the standard 
model are what they are in the model, and not only \textsl{what} are they 
and \textsl{how} do they work. In other words, the question is: 
\textsl{Is it necessarily so??}

Another reason (more pragmatic) is that often in mathematics, questions 
originating in Nature tend to be more seminal than others imagined ``out of 
the blue". That is, if I may say so, an experimental fact. It is therefore 
worth developing the new mathematical tools we have indicated, and those that
their development will suggest. If that solves physical problems (those 
intended or others), so much the better. In any case that should give 
nontrivial mathematics.
We express the general approach as follows:   

\begin{conjecture}\textsc{The Deformation Conjecture}. 
{Internal symmetries of elementary particles emerge from their relativistic 
counterparts by some form of deformation (possibly generalized, including 
quantization).}
\end{conjecture}

In particular we would like to realize a \textsl{Quantum Deformation Dream}:
\begin{itemize}
\item
The above mentioned ``Quantum Deformations" can be defined, then permit 
to define ``QQgroups," 
including a ``restricted" version thereof (at roots of unity) which would be 
finite-dimensional algebras, and the tensor products of their representations 
can be studied.
\item
Such a procedure can be applied to $\mathcal{U}(\mathfrak{so}(3,2))$
(and $\mathcal{U}(\mathfrak{so}(1,2))$ as a toy model), if needed along 
with a supersymmetric extension and maybe some kind of ``affinization", 
and serve as a starting point for a well based theory of strong interactions.  
\end{itemize}
Of course one cannot rule out that symmetries, one of the bases of 
``the unreasonable effectiveness of mathematics in the natural sciences"
\cite{Wi60}, turn out to have little role in particle physics. I strongly 
doubt it.

In the list of problems we have encountered, the ``simplest" on the 
mathematical side seems to be what happens with tensor products of 
representations of qAdS (or of other rank 2 quantum groups) at root of unity. 
In a similar spirit it would be possibly more interesting to study 
representations of multiparameter deformations of AdS, including their
tensor products. 

After some mathematical and theoretical progress we should then try and see 
how that knowledge can be used to interpret the known strong interactions, 
step by step, starting with the earliest known particles. That requires 
both theoretical and phenomenological studies, and possibly also renewed 
experiments to check the theoretical results.

A more appealing approach is to try and develop a ``quantum
deformations" theory, in particular with quaternions. Then, if and when we
have such a theory, would come the question to apply it to AdS both in
mathematics and in possible theoretical physics applications. 

Ultimately one would have to study in details the phenomenological implications
of these developments. If they differ somewhat from the present interpretation
of the raw experimental data, we would need to revise that interpretation and 
possibly to re-do some experiments. 

Note that a ``fringe benefit" of any such 
revision is that it can be done using the present experimental tools, 
if needed (as far as more ancient data are concerned) with the refinement of 
modern technology. The civil society is not likely to give us significantly
more powerful accelerators.  

All these are problems worthy of attack. It can be expected that they will
prove their worth by hitting back. 
Starting from a primary question, I have asked many more, 
combined that with many notions and results developed during half a century
of research in physical mathematics, and indicated avenues along which some 
of these might be answered. At 75 I leave it to the next generations to 
enter that promised land and tackle the many problems that will follow.  


\subsection*{Acknowledgment}
This panorama benefited from 35 years of active collaboration with 
Moshe Flato (and coworkers),                                                   
from close to \verb+\infty+ discussions with many scientists
(too numerous to be listed individually) and from the hospitality of Japanese
colleagues during my so far over 10 years of relatively generous retirement 
from France, after 42 years of (sometimes bumpy) career in CNRS. 
I want also to thank the Editors of this volume for their friendly patience.

\end{document}